\documentclass[aps,prl,showpacs,%
preprint,
preprintnumbers,superscriptaddress,nofootinbib,english]{revtex4}
\usepackage{graphicx}
\usepackage{epstopdf}

\providecommand{\neel}{CNRS-N\'{e}el, 25 Avenue des Martyrs, 
38042 Grenoble cedex 9, France}
\providecommand{\csnsm}{CSNSM,
IN2P3-CNRS, Universit\'e Paris XI, bat 108, 91405 Orsay,  France}
\providecommand{\iek}{Universit\"{a}t Karlsruhe (TH), 
Institut f\"{u}r Experimentelle Kernphysik, Gaedestr. 1, 76128 Karlsruhe, Germany}
\providecommand{\fzk}{Forschungszentrum Karlsruhe, Institut f\"ur Kernphysik, 
Postfach 3640, 76021 Karlsruhe, Germany}
\providecommand{\iramis}{CEA, Centre d'Etudes Saclay, 
IRAMIS, 91191 Gif-Sur-Yvette Cedex, France}
\providecommand{\irfu}{CEA, Centre d'Etudes Saclay, IRFU, 
91191 Gif-Sur-Yvette Cedex, France}
\providecommand{\ipnl}{IPNL, Universit\'{e} de Lyon, Universit\'{e} Lyon 1, 
CNRS/IN2P3, 4 rue E. Fermi 69622 Villeurbanne cedex, France}
\providecommand{\jinr}{Laboratory of Nuclear Problems, JINR, Joliot-Curie 6, 
141980 Dubna, Moscow region, Russia}

\usepackage{babel}
\begin{document}

\preprint{Draft 04/05/09}

\title{A new high-background-rejection dark matter Ge cryogenic detector}

\author{The EDELWEISS Collaboration\\A.~Broniatowski}\affiliation{\csnsm}
\author{X.~Defay}\affiliation{\csnsm}
\author{E.~Armengaud}\affiliation{\irfu}
\author{L.~Berg\'e}\affiliation{\csnsm}
\author{A.~Benoit}\affiliation{\neel}
\author{O.~Besida}\affiliation{\irfu}
\author{J.~Bl$\mbox{\"u}$mer}\affiliation{\iek}\affiliation{\fzk}
\author{A.~Chantelauze}\affiliation{\fzk}
\author{M.~Chapellier}\affiliation{\csnsm}
\author{G.~Chardin}\affiliation{\csnsm}
\author{F.~Charlieux}\affiliation{\ipnl}
\author{S.~Collin}\affiliation{\csnsm}
\author{O.~Crauste}\affiliation{\csnsm}
\author{M.~De~Jesus}\affiliation{\ipnl}
\author{P.~Di~Stefano}\affiliation{\ipnl}
\author{Y.~Dolgorouki}\affiliation{\csnsm}
\author{J.~Domange}\affiliation{\csnsm}
\author{L.~Dumoulin}\affiliation{\csnsm}
\author{K.~Eitel}\affiliation{\fzk}
\author{J.~Gascon}\affiliation{\ipnl}
\author{G.~Gerbier}\affiliation{\irfu}
\author{M.~Gros}\affiliation{\irfu}
\author{M.~Hannawald}\affiliation{\irfu}
\author{S.~Herv\mbox{\'e}}\affiliation{\irfu}
\author{A.~Juillard}\affiliation{\ipnl}\affiliation{\csnsm}
\author{H.~Kluck}\affiliation{\fzk}
\author{V.~Kozlov}\affiliation{\fzk}
\author{R.~Lemrani}\affiliation{\irfu}
\author{A.~Lubashevskiy}\affiliation{\jinr}
\author{C.~Marrache}\affiliation{\csnsm}
\author{S.~Marnieros}\affiliation{\csnsm}
\author{X-F.~Navick}\affiliation{\irfu}
\author{C.~Nones}\affiliation{\csnsm}
\author{E.~Olivieri}\affiliation{\csnsm}
\author{P.~Pari}\affiliation{\iramis}
\author{B.~Paul}\affiliation{\irfu}
\author{S.~Rozov}\affiliation{\jinr}
\author{V.~Sanglard}\affiliation{\ipnl}
\author{S.~Scorza}\affiliation{\ipnl}
\author{S.~Semikh}\affiliation{\jinr}
\author{M-A.~Verdier}\affiliation{\ipnl}
\author{L.~Vagneron}\affiliation{\ipnl}
\author{E.~Yakushev}\affiliation{\jinr}

\begin{abstract}
A new design of a cryogenic germanium detector for dark matter search is presented, 
taking advantage of the coplanar grid technique of event localisation for improved 
background discrimination. 
Experiments performed with prototype devices in the EDELWEISS II setup at the 
Modane underground facility demonstrate the remarkably high efficiency of these
devices for the rejection of low-energy $\beta$, approaching 10$^5$ . 
This opens the road to investigate the range beyond 10$^{-8}$ pb in 
the WIMP-nucleon collision cross-sections, as proposed in the EURECA 
project of a one-ton cryogenic detector mass.

\end{abstract}


\maketitle

Dark Matter embodies one of the outstanding enigmas of contemporary physics, 
for which supersymmetric (SUSY) theories propose a solution in the form of 
non-baryonic particles, the WIMPs (Weakly Interacting Massive Particles)~\cite{rev}. 
WIMPs are predicted to have exceedingly low interaction rates with baryons, 
which requires a highly efficient background rejection. 
To this end, the EDELWEISS~\cite{edw} and the CDMS~\cite{cdms} experiments 
rely on germanium cryogenic detectors, which provide such an efficient background 
discrimination, with  however limited efficiency for surface interactions~\cite{shutt}. 
Methods have been devised to identify such surface interactions,
based on the use of non-equilibrium phonon sensors~\cite{akerib,marnieros}. 
This letter describes an alternative solution to this problem, which is
at the same time simple and flexible.
It is based on the coplanar grid technique 
for event localisation~\cite{luke-ieee,amman,brink}. 
Measurements in low background conditions at LSM (Laboratoire Souterrain de Modane) 
demonstrate the efficiency of these devices to discriminate nuclear recoil events, 
as expected from WIMP elastic scattering, against both the $\gamma$ and $\beta$ 
backgrounds of the experiment. 
Evidence is thus given that detector sensitivities better than  10$^{-8}$ pb in the 
WIMP nucleon cross-sections can be achieved, representing an improvement by more 
than a factor of 5 over the present limits~\cite{cdms,xenon}, and sampling a large part of the 
SUSY parameter space without appreciable background contamination. 

The latest data from the CDMS~\cite{cdms} and the XENON\cite{xenon} collaborations 
set upper limits for the WIMP-nucleon cross-sections of about 5$\times$10$^{-8}$ pb 
for a WIMP mass in the range of 20 up to 100 GeV. 
In germanium detectors, this amounts to a count rate not exceeding 5$\times$10$^{-3}$ 
event/kg/day for deposited energies between 15 keV and 65 keV. 
For comparison, the average $\gamma$ rate within the shields of the EDELWEISS II 
experiment at LSM is about 10 events/kg/day in the same energy range. 
A $\gamma$ background rejection better than 1 in 10$^{4}$ is therefore needed 
to achieve a 10$^{-8}$ pb sensitivity in WIMP-nucleon interactions. 
The strategy for $\gamma$ background discrimination is based on the fact that a 
nuclear recoil event in germanium has a reduced ionisation yield 
(typically by about a factor of 3), as compared with an electron recoil triggered 
by ionising radiation. 
A highly efficient method for $\gamma$ background rejection follows, 
based on the dual measurement of the ionisation charge and the energy deposited 
in the form of heat in the detector~\cite{shutt2}. 
An additional background remains, however, due to a minute contamination of 
the detector surfaces and surroundings with $^{210}$Pb~\cite{fiorucci}. 
A daughter of Rn, $^{210}$Pb emits X-rays and conversion electrons with energies 
falling precisely within the energy range of interest.  
It further decays to $^{210}$Bi, which is a $\beta$ emitter with an end-point 
at 1.16 MeV, and to $^{210}$Po, an $\alpha$ emitter. 
The rate of low-energy ($<60$ keV) electron impacts
is about 10$^{-2}$ /cm$^2$/day, 
which amounts to approximately 2 events per kg of fiducial mass and per day
and requires a rejection better than 1 in 2000. 
This contribution remained partially unresolved in the detectors previously used by 
EDELWEISS, due to their poor charge collection characteristics 
for surface energy deposits~\cite{shutt}. 
The resulting pulse height defect entails a confusion between the $\beta$ background, 
and the nuclear recoils expected from WIMPs. 
Initially proposed by the CDMS collaboration~\cite{brink}, 
the method we have developed to resolve this issue is essentially a variation 
of the coplanar grid technique~\cite{luke-ieee,amman}, 
in which interleaved strips are substituted for the classical disk-shaped collection electrodes. 
The depth of an event relative to the surfaces can be inferred from a comparison 
of the ionisation signals on the different strips, making possible a rejection of energy 
deposits at the detector surfaces~\cite{defaythese,broniatowski-2008,defay}.

Figure 1 (a) is a schematics of a 200 g prototype detector, 
tested at LSM together with two other 400 g detectors of a similar design. 
Following a surface passivation treatment~\cite{shutt}, 
aluminum electrodes 250 nm thick were evaporated onto the top and the bottom 
surfaces of the Ge crystal, in the form of annular concentric rings 200 $\mu$m 
wide with a 2 mm pitch. 
The rings are alternately connected by ultra-sonic bonding, yielding four sets of 
interleaved electrodes: $a$, $b$, $c$ and $d$ 
at potentials $Va$, $Vb$, $Vc$ and $Vd$ respectively, 
each fitted with a charge-sensitive amplifier. 
Guard electrodes $g$ and $h$ at potentials $Vg$ and $Vh$ 
(also fitted with charge amplifiers), 
and a neutron transmutation-doped (NTD) Ge thermometer complete the device. 
A calculation of the field geometry leads to divide the volume of the detector 
into different regions of charge collection, as defined by the dotted lines in the figure. 
Events in the bulk (region (1)) deliver charges to the $b$ and $d$ sets of electrodes on 
opposite sides of the detector, 
and those in the guard region (2) to the $g$ and $h$ electrodes respectively. 
On the other hand, surface and near-surface events (regions (3)) 
deliver charges to electrodes on the same side of the detector 
($a$ and $b$, or $c$ and $d$ respectively). 
By applying appropriate selection cuts in the signal amplitudes 
in the channels $a$, $c$, $g$ and $h$ (called thereafter the veto channels), 
bulk events are retained alone, whereas the guard, the surface and 
near-surface events are eliminated. 
Events where energy is deposited along the dotted lines exhibit charge 
sharing between more than two measurement channels and are also rejected 
by these selection rules. 
Figure 1 (b) is a map of the field in the area close to the bottom surface. 
Note, in particular, {\em i)} the enhancement of the field under the electrodes 
($>$ 10V/cm) compared to the average in the bulk 
(approximately 0.6 V/cm for the voltage biases in the figure), and 
{\em ii)} the occurrence of low-field areas ($<$ 0.2 V/cm) 
at some depth below the surface (2 mm typically), 
surrounding singular points of field cancellation~\cite{kozorezov}. 
We first present the overall performance of the detector with respect to 
$\beta$ and $\gamma$ background rejection. 
We then analyse more specifically the detector response to energy 
deposits at the surfaces.  

In order to study with high statistics the $\beta$ background of the experiment, 
the 200 g detector was fitted with two $^{210}$Pb sources 
(obtained by Rn implantation in a copper substrate), 
each facing one of the detector surfaces. 
Scatter plots of the ionisation yield versus the deposited energy are obtained 
from the amount of heat evolved by an event, and from the amplitudes of the charge 
signals induced in the different measurement channels~\cite{defaythese,broniatowski-2008} 
(this includes a correction for the Joule heat produced in the course of charge collection, the Neganov-Luke effect~\cite{neganov,luke}). 
Figure 2 presents the scatter plots obtained (a) before, and (b) after surface event rejection.
The scatter plot (b) was deduced from (a) by applying selection cuts in the charge signal 
amplitudes, any event with an amplitude in the veto channels larger than 
2 keV -- twice the standard deviation of the ionisation baseline noise -- being rejected. 
Out of a total of 120,000 $\beta$ events recorded above 15 keV 
and with an ionisation yield below 0.6, only 3 near the energy threshold 
survive the selection cuts. Only one of them falls within the nuclear recoil band. 
This correspond to the leaking into that band of one low-energy
electron from $^{210}$Pb decay out of an incident flux of approximately
60,000, thus exceeding by a factor 30 the rejection required for
a 1000 kgd experiment.

Further tests were made using a $^{133}$Ba (356 keV) $\gamma$ source 
in order to check the detector response to electron recoils in the bulk. 
Following selection cuts, out of a total of 10$^{5}$  (10$^4$) events in the energy 
range from 15 to 400 keV (15 to 65 keV), none have an ionization yield inferior to 0.5.

A last issue to address is the magnitude of the fiducial mass left after the selection 
cuts were made. 
Figure 2 (b) shows the 10 keV K line from the cosmogenic $^{68}$Ge and 
$^{65}$Zn isotopes, homogeneously distributed within the volume of the detector. 
The intensity of this line before and after selection can be used to determine 
the fiducial mass. 
This was done with the two 400 g detectors, yielding a value of 160 g per detector, 
in agreement with an estimate from the field map, and primarily limited by the 
guard regions (figure 1). 
The same result was also obtained by a neutron calibration using an Am-Be source. 

The specifications for a 10$^{-8}$ pb dark matter experiment are thus met 
both in terms of the $\beta$ and $\gamma$ background rejection capabilities 
of the detectors.   

Further experimental and modelling studies were made to analyze the 
detector response to energy deposits at the surfaces, using $^{109}$Cd 
sources (62 and 84 keV electrons and 88 keV photons). 
Surface events are identified unambiguously from their reduced ionisation 
yield compared to the more penetrating $\gamma$Õs. 
They are thus found of two different kinds: the vast majority (over 90\%, 
exclusive of those underneath the guards) 
deliver charges to two measurement channels on the same side of the detector, 
one of them a veto channel (these are just a special case of energy deposits 
in region (3), fig. 1). 
The remaining fraction shows charge sharing between three channels, 
one of them a veto again. 
The proportion of events in both categories scales approximately as the relative 
areas in-between and underneath the $b$ and $d$ electrodes 
(with a ratio of 20 to 1 in the geometry of Fig. 1), 
which strongly suggests that 3-channel events are associated to 
energy deposits across these electrodes. 

To help understand these features of the detector signals, 
ionisation pulses were digitized using a 
wide-band electronics, and compared to those obtained from a computer 
modelling of the charge collection~\cite{broniatowski-2004}.
The model includes as salient features
{\em i)} the carrier Coulomb interactions, an essential factor to account for 
cloud expansion and charge sharing between different measurement 
channels~\cite{gatti}, 
{\em ii)} hot carrier effects in diffusion and drift at cryogenic 
temperatures~\cite{sunqvist,aubry} and 
{\em iii)} a simple prescription for carrier trapping and recombination 
at the surfaces, namely that any carrier drifted onto a free surface 
(resp., a metal electrode) becomes trapped 
(resp., recombines with a carrier of the opposite sign). 
Based on this model, computer simulation provides the precise time 
structure of the ionization signals for the charge sharing events. 
Figure 3 (a) shows, as a typical example, time-resolved recordings of the 
charge signals for an electron impact 
from a $^{109}$Cd electron source 
on one of the $d$ strips 
(geometrical configuration shown in fig. 3 (c)). 
The shape of the recordings was compared to a library of simulated signals 
for a set of electron impacts at different locations across the width of the strip. 
Figure 3 (b) represents the best fit simulation for the event under consideration, 
corresponding to energy deposition close to mid-width of the electrode, 
and reproducing the finer details of the recorded signals. 
Examination of the carrier trajectories (fig. 3 (c)) gives insight into the physical 
effects involved in surface event discrimination. 
The dominant feature is cloud expansion, which causes the carriers to scatter 
away from the field line passing through the point of energy deposition. 
The scatter is further enhanced due to the inhomogeneity of the field 
underneath the electrodes (fig. 1 (b)), so that a fraction of the carriers is 
collected at a veto electrode. 
A similar argument applies to the energy deposits within the 
low-field areas (fig. 1 (b)), as the cloud expansion is instrumental here again 
in driving a fraction of the collected charges towards a veto channel, 
which permits to reject this category of events as well~\cite{broniatowski-2008}.

Besides proving highly effective for both $\beta$ and $\gamma$ background rejection, 
major assets of these new detectors are their ease of fabrication and basic simplicity 
of operation and data analysis. 
The program of the EDELWEISS collaboration aims to increase the cumulated 
mass of the detectors in operation at LSM, 
so that the physically significant 10$^{-8}$ pb range of spin-independent 
WIMP-nucleon cross-sections can be probed in a reasonably short period of time. 
In accordance with this objective, effort in R\&D is devoted to maximizing 
the fiducial volume of the detectors through the use of modified electrode 
geometries, increasing the detector masses up to the kg scale, 
and reducing the production costs by resorting to lower purity grade 
germanium crystals. 
Looking ahead of these developments, Ge coplanar grid detectors of a 
suitable design are thus one of the promising candidates for the EURECA 
project of a one-ton detector dark matter experiment~\cite{eureca}.

\begin{acknowledgments}
The help of the technical staff of the Laboratoire 
Souterrain de Modane and the participant laboratories is 
gratefully acknowledged. 
This project is supported in part by Agence Nationale pour la Recherche 
under contract ANR-06-BLAN-0376-01.
\end{acknowledgments}

\newpage

\begin{figure}
\begin{center}
\end{center}
\caption{(a): Detector in cross-section, showing the different sets of collection electrodes (see text). 
The electrodes are enlarged for better visibility. 
The detector is in the form of a cylinder 48 mm in diameter and 20 mm in height, 
with axial symmetry along the vertical $z$ axis. 
Voltage biases: $Va$ = -0.75 V, $Vb$ = +2.0 V, 
                             $Vc$ = 0.75 V, $Vd$ = -2.0 V, 
                             $Vg$ = +0.5 V, and $Vh$ = -0.5V. 
The dotted lines delimit the respective areas of charge collection, namely 
(1) the bulk (or fiducial) volume; 
(2) the guard, and 
(3) the near-surface areas of the detector, respectively. 
The zero-field points are marked by a solid dot. 
Field lines through points of energy deposition are drawn for 
representative events in areas (1) and (2), 
marked as o and * respectively; 
(b) enlarged field map of the area above the bottom surface of the detector, 
showing a representative event (+) in area (3) with the corresponding field line. 
Contours of equal field strength are marked as the dot-dashed lines. 
$r$ and $z$ are the radial and the vertical coordinates, respectively.}
\end{figure}

\begin{figure}
\caption{Scatter plots of the ionisation yield versus the recoil energy recorded 
in a 200 g prototype detector exposed to $^{210}$Pb sources, 
(a) before and (b) after applying the cuts to reject near-surface events. 
Bias voltages are the same as in figure 1. 
The ionisation yield is normalised to unity for electron recoil 
events with full charge collection. 
The total number of events in (a) is 185,000, 
out of which 50,000 are due to the 5.3 MeV alphas (labelled $\alpha$) 
and 11,000 are due to the 46 keV transition from $^{210}$Pb ($\gamma$). 
The rest consists essentially of the $\beta$ events (120,000) 
from the radioactive sources. 
After the cuts, the remaining events originate from the gamma background 
of the experiment and from the 10 keV (K) line of the cosmogenic $^{68}$Ge 
and $^{65}$Zn isotopes. 
The experimental threshold (dashed lines) is determined from the thresholds 
of the ionisation and heat signals (4 and 8 keV respectively). 
90\% of the nuclear recoils are expected between 
the two dotted lines in panel (b).}
\end{figure}

\begin{figure}
\caption{(a) Time-resolved recordings of the ionisation signals for an 84 keV 
electron impact ($^{109}$Cd source) on a $d$ electrode. 
The signals in the $b$, $c$ and $d$ channels are represented by the dotted, 
the dashed, and the full lines respectively. 
Amplitudes are normalised to unity for the charge collected in the $d$ 
measurement channel. 
Electrode biases: $Va$ = 0 V, $Vb$ =  +2.5 V, $Vc$ =  0 V, $Vd$ = - 2.5 V, 
$Vg$ =  +0.5 V, and $Vh$ = - 0.5V, respectively; 
(b) best fit simulation of an electron impact on a $d$ electrode (see text);
(c) electron trajectories for the event simulated in (b). 
The field line through the point of energy deposition is represented by the thick line. 
The wavy form of the trajectories is an effect of carrier diffusion. 
Due to cloud expansion, electrons share between 
the $b$ and $c$ electrodes on opposite sides of the detector. }
\end{figure}

\clearpage

Figure 1 a
\begin{center} \includegraphics[scale=0.5]{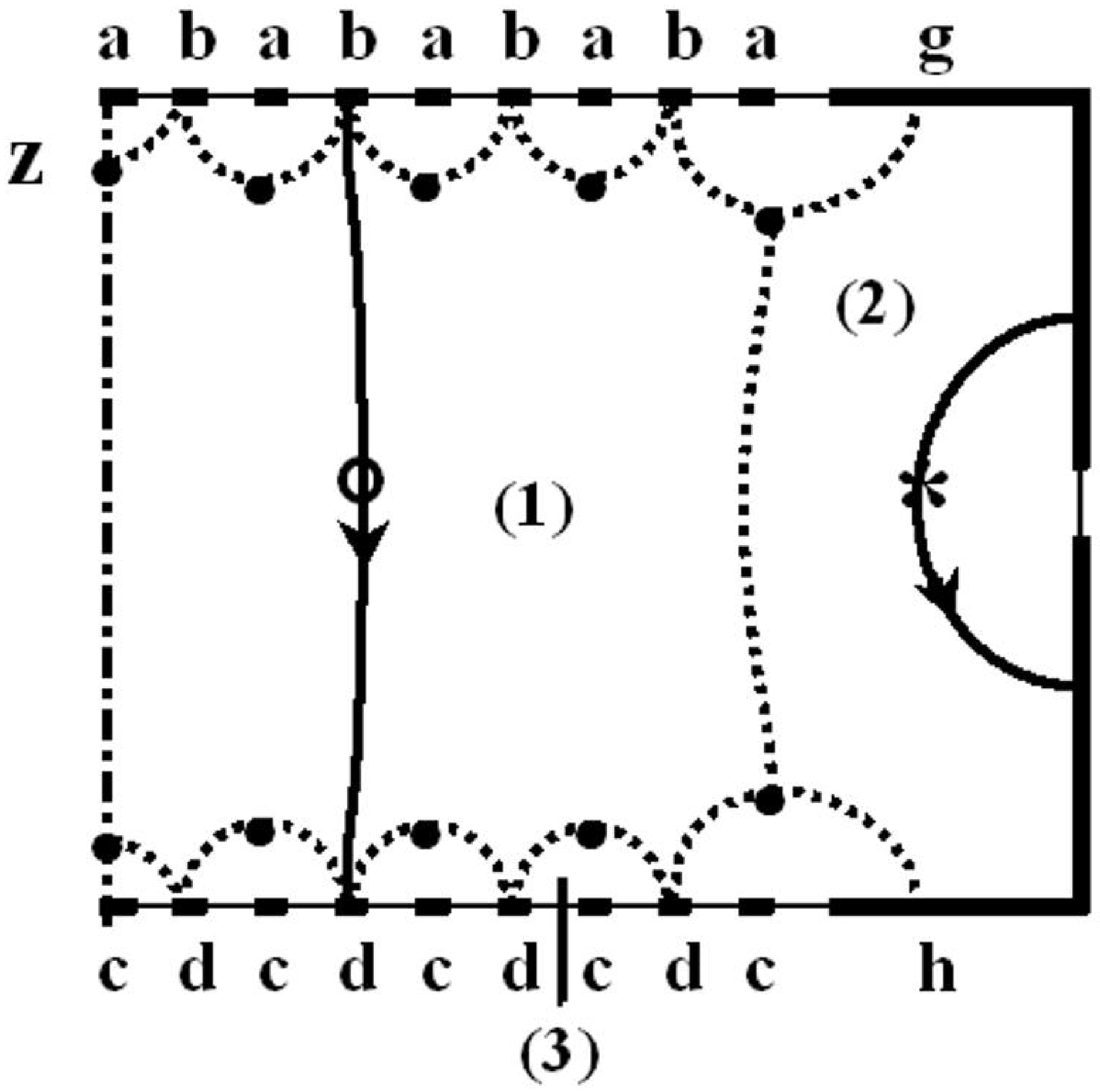} \end{center}

Figure 1 b
\begin{center} \includegraphics[scale=0.4]{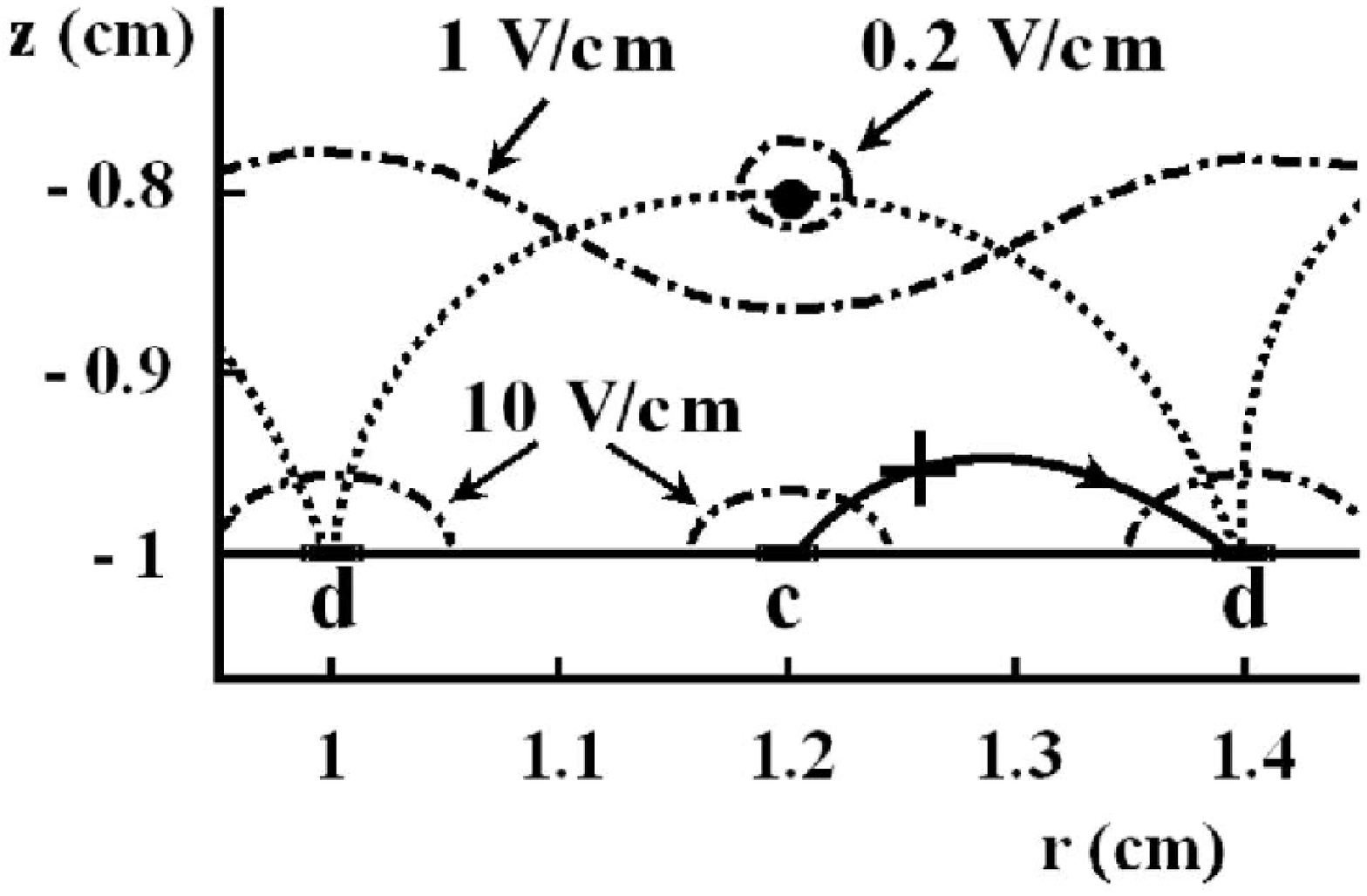} \end{center}

\clearpage

Figure 2
\begin{center}\includegraphics[scale=0.75]{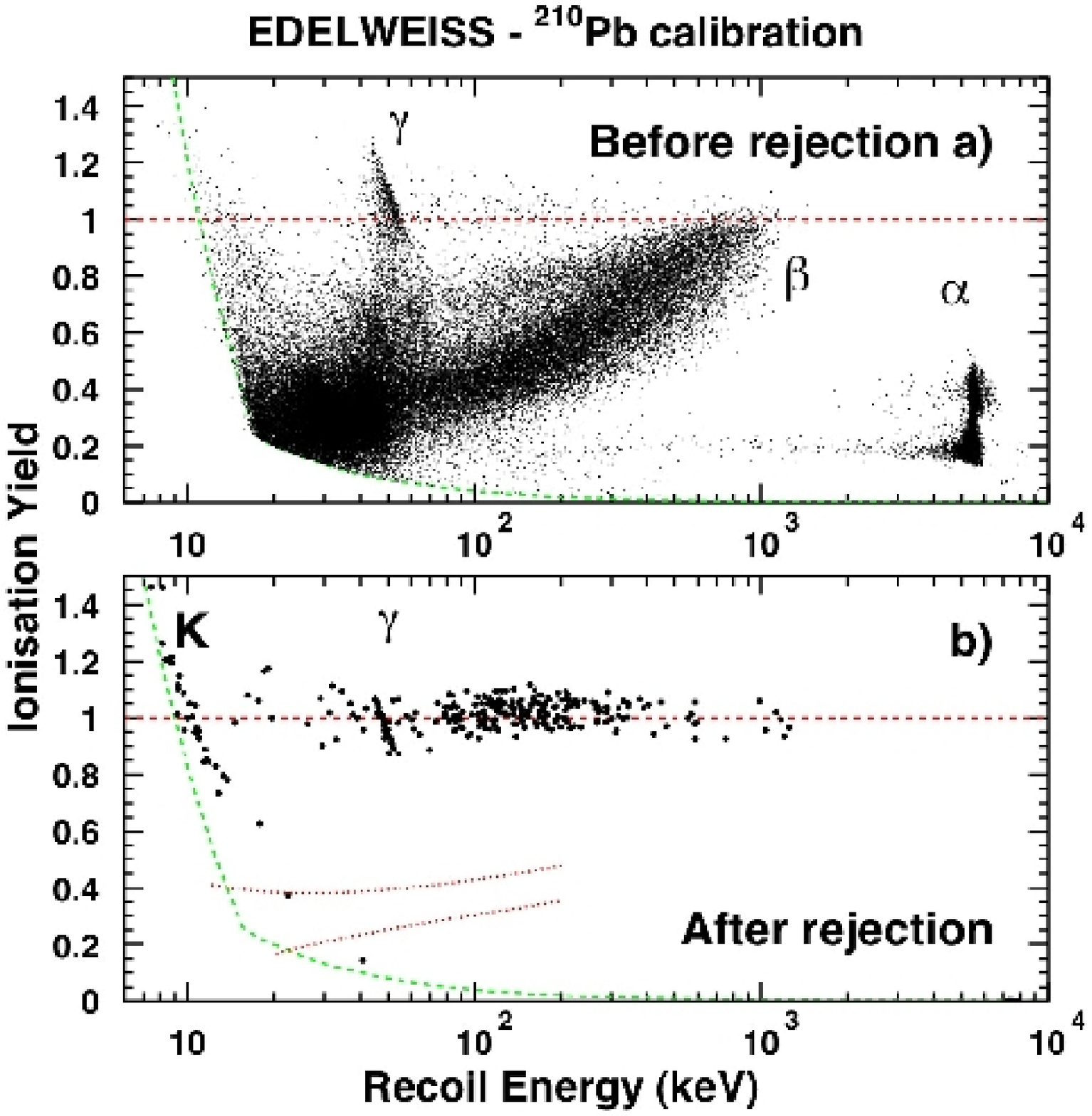}\end{center}
\clearpage

Figure 3
\begin{center} \includegraphics[scale=0.53]{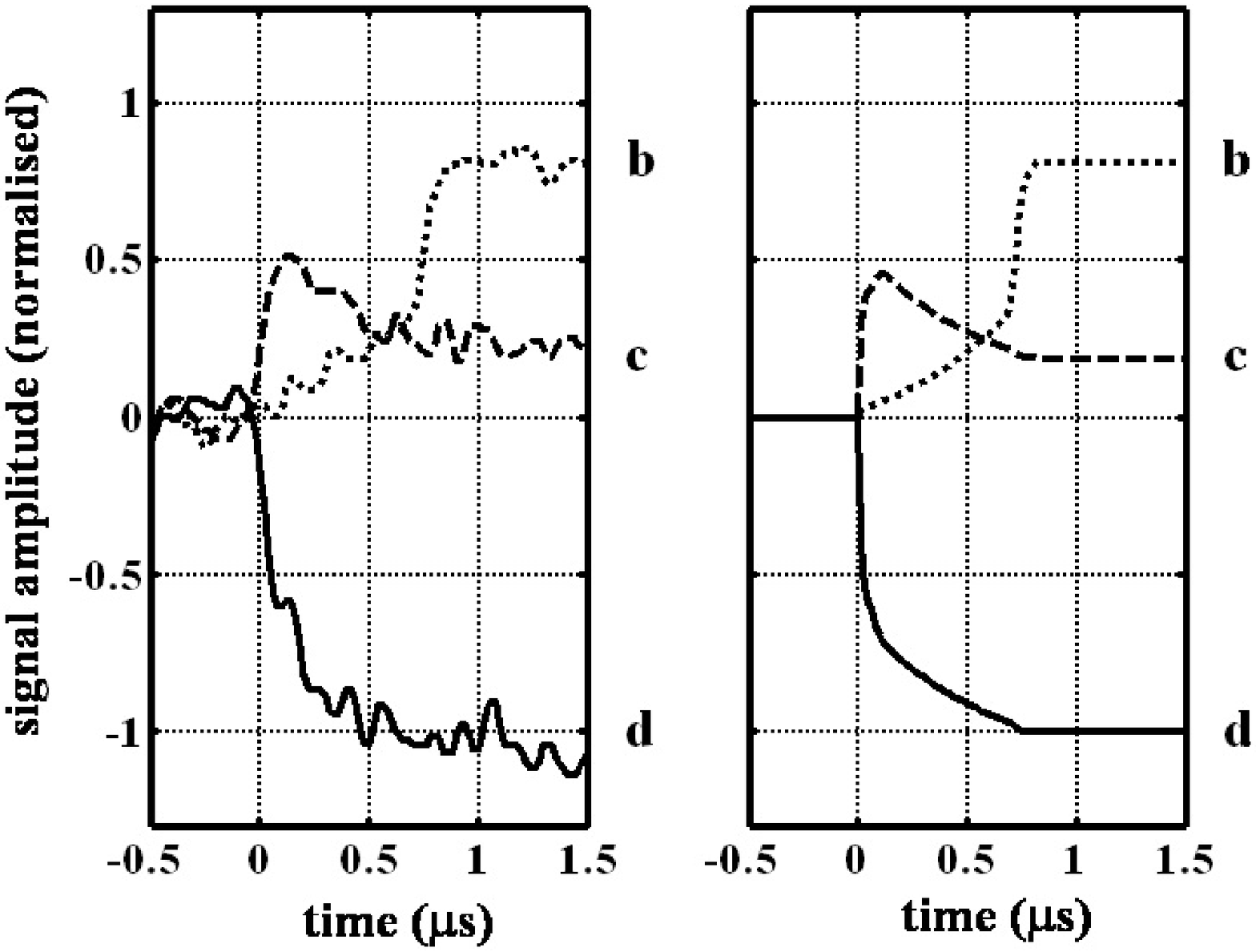}%
\includegraphics[scale=0.3]{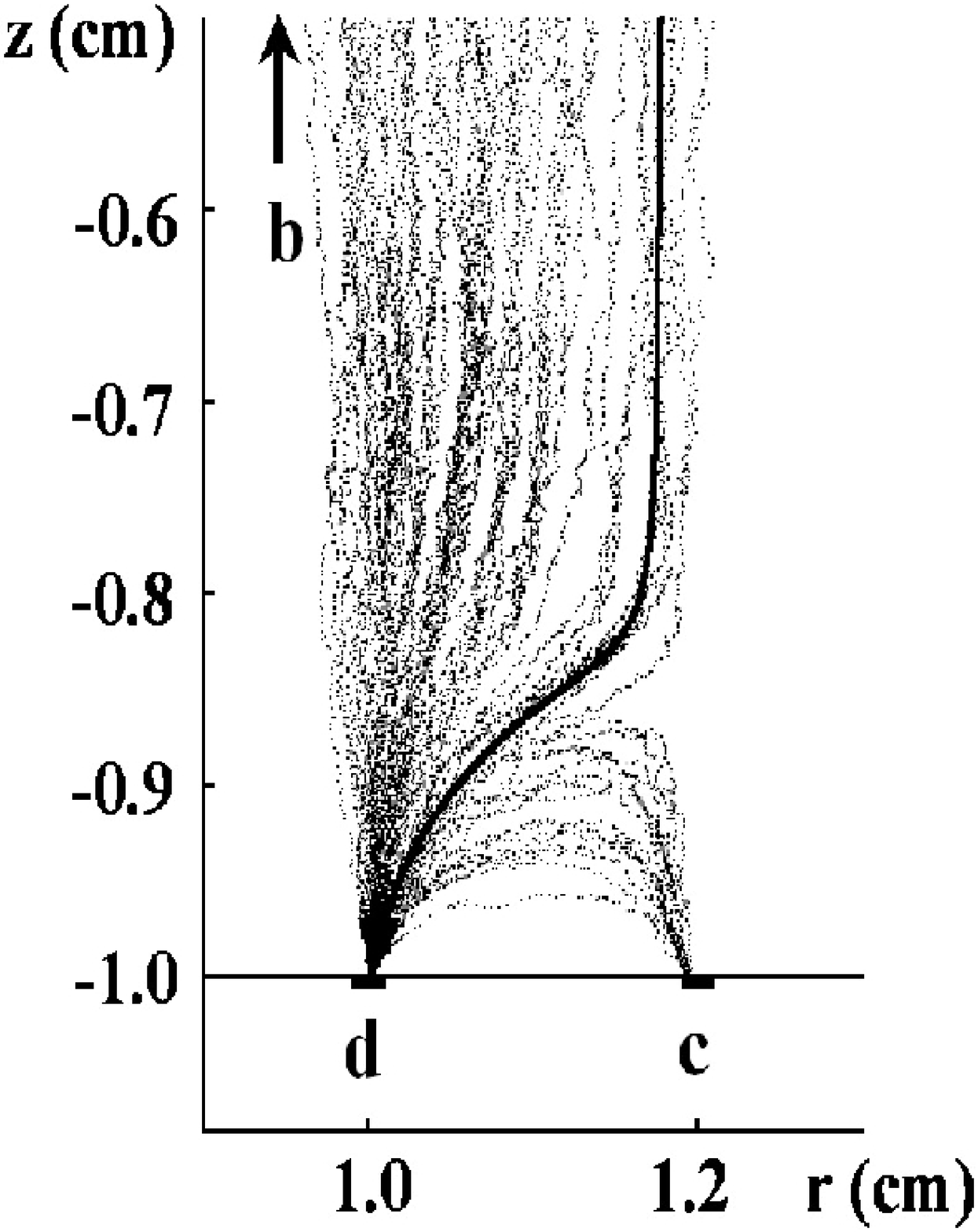}
\end{center}
\begin{center}
\large{\bf{ a) \hspace{5.0cm} b) \hspace{5.0cm} c)}}
\end{center}

\end{document}